\documentclass[superscriptaddress, amsmath,amssymb, pre,twocolumn,longbibliography]{revtex4-1}

\usepackage{graphicx}
\usepackage{dcolumn}
\usepackage{bm}
\usepackage{hyperref}
\usepackage{color}
\usepackage{setspace}

\newcommand{\angstrom}{\textup{\AA}}

\begin{document}

\title{Non-Hookean statistical mechanics of clamped graphene ribbons}

\author{Mark J.~Bowick}
\affiliation{Department of Physics, Syracuse University, Syracuse, NY 13244, USA}
\author{Andrej Ko\v smrlj}%
\affiliation{Department of Mechanical and Aerospace Engineering, and Princeton Institute for the Science and Technology of Materials (PRISM), Princeton University, Princeton, NJ 08544, USA}%
\author{David R.~Nelson}%
\affiliation{Department of Physics, Department of Molecular and Cellular Biology and School of Engineering and Applied Sciences, Harvard University, Cambridge, MA 02138, USA}%
\author{Rastko Sknepnek}%
\affiliation{School of Science and Engineering and School of Life Sciences, University of Dundee, Dundee, DD1 5EH, United Kingdom}%

\date{\today}

\begin{abstract}
Thermally fluctuating sheets and ribbons provide an intriguing forum in which to investigate strong violations of Hooke's Law:  large distance elastic parameters are in fact not constant, but instead depend on the macroscopic dimensions. Inspired by recent experiments on free-standing graphene cantilevers, we combine the statistical mechanics of thin elastic plates and large-scale numerical simulations to investigate the thermal renormalization of the bending rigidity of graphene ribbons clamped at one end. For ribbons of dimensions $W\times L$ (with $L\geq W$), the macroscopic bending rigidity $\kappa_R$ determined from cantilever deformations is independent of the width when  $W<\ell_\textrm{th}$, where $\ell_\textrm{th}$ is a thermal length scale, as expected. When $W>\ell_\textrm{th}$, however, this thermally renormalized bending rigidity begins to systematically increase, in agreement with the scaling theory, although in our simulations we were not quite able to reach the system sizes necessary to determine the fully developed power law dependence on $W$. When the ribbon length $L > \ell_p$, where $\ell_p$ is the $W$-dependent thermally renormalized ribbon persistence length, we observe a scaling collapse and the beginnings of large scale random walk behavior.
\end{abstract}

\pacs{Valid PACS appear here}
\maketitle


\section{Introduction}
\label{sec:intro}

Two dimensional crystals as mechanical objects are, at first glance, rare and delicate, but graphene belies expectations. It is a robust 2D membrane with unique and tunable material properties that can be exploited in micro- and nanoscale metamaterials.  While the electronic properties of graphene are well understood, the continuum mechanical behavior remains a frontier of research. In particular, experimental studies on graphene~\cite{blees15,nicholl15}, along with previous theory~\cite{nelson87, aronovitz88,guitter88,aronovitz89,guitter89,ledoussal92,bowick01PhysReports,bowick09,gazit09,kownacki09,zakharchenko10,braghin10,hasselmann11,kosmrlj16} and Monte Carlo simulations~\cite{zhang93,bowick96,bowick97,los09,roldan11,paulose12,troster13,troster15,los16}, indicate that thermal effects \emph{dramatically modify} the mechanical properties of the membrane (see also Refs.~\cite {nelsonB, katsnelsonB, wieseB,amorim16}). Here we study how the interplay between thermal effects and boundary conditions, as well as geometry, affect the measured mechanical properties of graphene sheets. The results presented here are also directly applicable to the study of the thermalized behavior of other free-standing covalently-bonded atomically thin membranes, such as $\rm{MoS}_2$~\cite{wang12}. 
The extended translational order typical of crystals can be unstable to thermal fluctuations, and the situation is even more precarious when the crystalline sheet is a membrane free to fluctuate in the third dimension. Although thermal fluctuations will eventually decorrelate long range order in the membrane normals in liquid membranes (such as lipid bilayers), for crystalline membranes the nonlinear coupling of height fluctuations to in-plane phonon deformations leads to a length-scale dependent stiffening of the microscopic bending rigidity at long length scales~\cite{nelson87}.

To emphasize the remarkable nature of the ordered phase of crystalline membranes note that if one views the local normals as classical vector spins then one would expect the model to be disordered in the same way as the 2d-Heisenberg model of magnetism, a consequence of the Mermin-Wagner-Hohenberg theorem~\cite{mermin66,hohenberg67}. A crystalline membrane differs from the Heisenberg model, however, in several crucial ways. First of all, the ``spins" are constrained because they must be strictly normal to a continuous underlying membrane surface. Such constraints, intimately connected with the existence of an in-plane shear modulus,  change the available configuration space and consequently alter the statistical mechanical behavior of the system, including the phase diagram itself. In a perturbative field theory treatment, these constraints lead to the nonlinear coupling of in-plane phonons to height that renormalize the bending rigidity so that it flows with length scale (or wavevector) rather than being a fixed material parameter at long wavelengths. Height fluctuations inevitably cost elastic energy because the two planar phonon degrees of freedom cannot compensate for the three degrees of freedom in the symmetric 2d strain tensor associated with an arbitrary height deformation.  The result is long-range flatness of the thermally stiffened membrane or, equivalently, long-range order in the normal-normal correlation function describing the spatial correlations of local normals to pieces of the membrane. While these effects have been studied theoretically suitable experimental systems have been more difficult to find.  

One might expect that thermal fluctuations of a membrane would only be important for extremely soft systems such as the spectrin cytoskeletal network of the red blood cell \cite{schmidt93}, spherical assemblies of spider silk proteins \cite{hermanson07} or polymersomes \cite{shum08}. Even here, however, the bending rigidity is typically much larger than the temperature at which such systems are stable and the corresponding length scales at which thermal fluctuations are important can be dozens of particle spacings or more. Graphene, however, has a very large Young's modulus $Y_0$ (order $20\mathrm{eV}\angstrom^{-2}$), but a relatively modest microscopic bending rigidity $\kappa_0 \approx 1.2$eV. Perturbative corrections to the microscopic bending  rigidity $\kappa_0$ due to thermal ripples for a fluctuating crystalline membrane are described by a scale-dependent bending rigidity of the form $\kappa_R(\mathbf{q}) = \kappa_0 + \frac{k_BTY_0}{\kappa_0}\mathcal{I}(\mathbf{q})$, where $\mathbf{q}$ is a wavevector, $\mathcal{I}$ is a momentum integral that scales as $1/q^2$ when $\mathbf{q} \to 0$, or equivalently as $L^2$, with $L$ being a long wavelength cutoff provided by the system size $L$ and $Y_0$ the 2D Young modulus \cite{nelson87, aronovitz88, ledoussal92, kownacki09, braghin10, nelsonB, bowick09}. One readily sees that corrections to the bare bending rigidity are of order the bare rigidity itself for $L>\ell_\textrm{th}$, with $\ell_\textrm{th}\approx\frac{\kappa_0}{\sqrt{k_BTY_0}}$. Remarkably, for graphene at room temperature this thermal length scale is extremely small, $\ell_\textrm{th} \approx 1.5\angstrom$, comparable to the spacing between carbon atoms.  The high in-plane elastic modulus of this covalently bonded material thus leads to significant thermal stiffening of the bending rigidity even at microscopic length scales! 

This shape stiffening is essential for stabilizing graphene as a 2d, approximately planar, crystal against the thermal fluctuations that often entropically dominate 2d systems with continuous symmetry. One should contrast covalently bonded sheets of graphene (or Mo${\rm S}_2$) with \emph{soft} matter,  where the much lower Young moduli mean that thermal fluctuations become strong only for much larger length scales. 

The correction to the bending rigidity, in units of $k_BT$, is proportional to $\textrm{vK}=\frac{Y_0 L^2}{\kappa_0}$, the dimensionless F{\"o}ppl-von K{\'a}rm{\'a}n number measuring the ratio of typical elastic deformation energies to bending energies \cite{lidmar03}. Large values of $\textrm{vK}$, even in the \emph{absence} of thermal fluctuations, lead to the notoriously difficult problems of thin plates and shells, important for understanding the strength of macroscopic objects such as domed sports arenas and submarines \cite{koiterB}. For an $L=200\mu\mathrm{m}$  square graphene sheet  the  F{\"o}ppl-von K{\'a}rm{\'a}n number is $\textrm{vK}\approx 10^{14}$, a number which can also be obtained by extrapolating the continuum elastic theory of thin plates of thickness $h$ as $vK\approx 10\left (\frac{L}{h}\right)^2$ \cite{lidmar03} for atomically thin graphene with $h\approx 1\angstrom$. To appreciate the enormous size of $\textrm{vK}$ in graphene, it is helpful to recall that the deformations involved in crumpling an ordinary piece of paper ``only" involve $\textrm{vK}\approx 10^6$. The $\textrm{vK}$ number in the thin limit is a predominantly geometric quantity determined by the aspect ratio of the material. Very large $\textrm{vK}$ numbers naively mean that bending should be a soft mode compared to elastic deformations. Here, however, the nonlinear coupling of height fluctuations to in-sheet phonons thermally stiffens the bending rigidity over scales larger than the microscopic mesh size and consequently stabilizes the extended crystalline sheet flat phase of graphene. This remarkable interplay of materials and concepts from both hard- and soft-matter physics is a striking feature of graphene and related materials, embodied in recent pioneering experiments on graphene ribbons by Blees, \emph{et al.}, who observed a $\sim\!4000-$fold enhancement of the bending rigidity at room temperature \cite{blees15}. In these experiments the $\sim10\mu\mathrm{m}$ wide ribbons were approximately $50,000$ times wider than they were thin.

In the rest of the paper we will utilize numerical simulations to measure the effect of thermal fluctuations on the bending rigidity of a clamped elastic ribbon and its dependence on the ribbon geometry. In Sec.~\ref{sec:numerical_model}, we describe the numerical method underlying our molecular dynamics simulations of thin ribbons clamped at one end along its width $W$. We find it convenient to use a triangular discretization of a elastic ribbon of length $L$, with a microscopic bending rigidity and Young's modulus adjusted to match the parameters of the dual honeycomb lattice of, say, a covalently bonded graphene sheet. Our numerical results are described and interpreted in terms of the renormalization group theory of ribbons in Sec.~\ref{sec:results}. We find a scale-dependent bending rigidity when $W$ exceeds the thermal length scale, and the beginnings of random walk behavior for ribbons when $L$ exceeds the thermally-renormalized persistence length. Concluding remarks appear in Sec.~\ref{sec:conclusions}. 

\begin{figure}[b]
\begin{centering}
\includegraphics[clip,width=0.95\columnwidth]{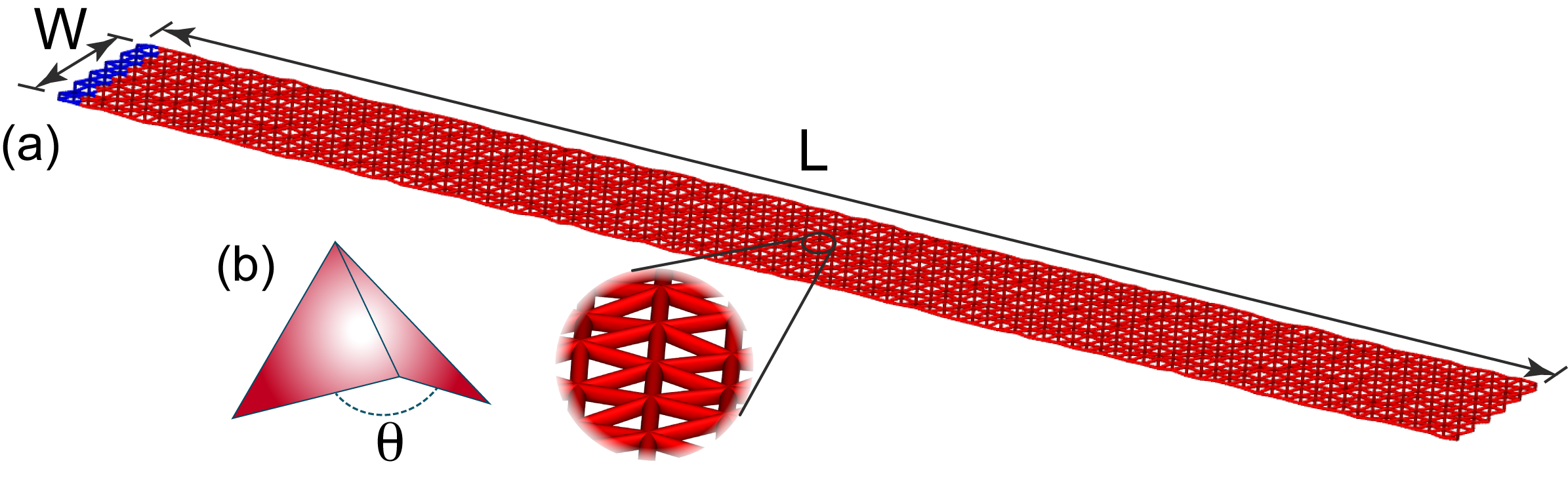}
\par\end{centering}
\caption{(Color online) (a) Our numerical simulations of a graphene strip of undeformed size $L\times W$ are preformed with a coarse-grained model commonly used to study elastic membranes \cite{seung88}. The strip is represented as an equilateral triangulation of a rectangle with bending and stretching energies defined along the edges and plaquettes of the triangulation. (b) For computational reasons, it is convenient to describe bending energy as a penalty of changing dihedral angle between two triangles sharing an edge. \label{fig:model}}
\end{figure}

\section{Numerical model}
\label{sec:numerical_model}

Instead of a fully atomistic description based on, say, the empirical bond-order (AIREBO) potential function \cite{stuart00,brenner02}, as used in Ref.~\cite{xu10} for example, or the approach of Ref.~\cite{zakharchenko09},  we have found it convenient to model a graphene strip using a coarse-grained dual representation commonly employed to study two-dimensional elastic membranes \cite{seung88}. With fixed computer resources, this strategy allows us significant gains in simulation sizes and speeds, without affecting the long time- and length-scale behavior we are studying here. The strip is discretized as a triangulation of a rectangle of size $L\times W$ (Fig.~\ref{fig:model}a). In the initial flat configuration, all triangles are equilateral with edge length $a$, which sets the microscopic length-scale in the model and will also be our unit of length.  A hexagon composed of six such triangles has the same symmetry as a graphene single crystal, and is assumed to be large enough to average over the detailed properties of a cluster of covalently bonded carbon atoms, but small enough not to affect the long time- and length-scale behavior of the macroscopic ribbon. The first two rows of vertices (in the $W$ direction, which we choose to coincide with the direction of the $y$ axis of the laboratory reference frame) remain immobile throughout the simulation. By fixing two rows of vertices we impose a boundary condition that fixes the normals and position of one ribbon edge, thus mimicking the clamping of one end of a strip in experiments in an otherwise unconstrained graphene experiment. The rest of the strip is free to move. The clamped region is included as part of the strip's initial (undeformed) length $L$. 

The elastic energy in our model calculations contains two terms, bending and stretching. The bending energy is described using a common discretization \cite{seung88} of the continuum bending energy:
\begin{equation}
E_{bend}=\frac{1}{2}\tilde{\kappa}\sum_{\left\langle IJ\right\rangle }\left|\mathbf{n}_{I}-\mathbf{n}_{J}\right|^{2},
\label{eq:bending}
\end{equation}
 where $\mathbf{n}_{I}$  is the unit-length normal to the triangle $I$ and the sum is carried out over all nearest neighbor pairs $\left\langle IJ\right\rangle$ of triangles. Triangle edges along the free sides and end of the strip are not treated in any special manner; if an edge is on the boundary it is assumed not to contribute to the bending energy. Eq.~(\ref{eq:bending}) can be rewritten as the dihedral energy associated with the edge $\left\langle IJ\right\rangle$  as
 \begin{equation}  E_{bend}=\tilde{\kappa}\sum_{\left\langle IJ\right\rangle }\left(1+\cos\theta_{IJ}\right),
  \label{eq:dihedral_bending}
 \end{equation}
where $\theta_{IJ}$  is the dihedral angle between two triangles sharing edge $\left\langle IJ\right\rangle$ (Fig.~\ref{fig:model}b). While the last two expressions are mathematically equivalent, computation of the dihedral forces is a standard feature of many molecular dynamics (MD) packages, thus allowing for a simple implementation in the existing MD software packages. The stretching energy is modeled by assigning harmonic springs of rest length $a$ and spring constant $\varepsilon$ to each edge \cite{seung88}, i.e.,
 \begin{equation}
 E_{stretch}=\frac{1}{2}\varepsilon\sum_{\left\langle i,j\right\rangle }\left(r_{ij}-a\right)^{2},
 \label{eq:stretching}
 \end{equation}
 where $r_{ij}=\left|\mathbf{r}_i-\mathbf{r}_j\right|$ is the Euclidean distance between nearest-neighbor vertices $i$ and $j$. Note that our discretization parameters $\tilde{\kappa}$ and $\varepsilon$ are directly related to the continuum Young's modulus, $Y_0=\frac{2}{\sqrt{3}}\varepsilon$ \cite{seung88} and bare continuum bending rigidity $\kappa_{0}=\frac{\sqrt{3}}{2}\tilde{\kappa}$ \cite{seung88,schmidt12}.

All numerical simulations were preformed using the HOOMD-blue molecular dynamics package \cite{anderson08} in the constant temperature (NVT) ensemble. The temperature was controlled using a  standard Nos\'e-Hoover thermostat \cite{nose84,hoover85} and was set to $T=1$. (In our simulations, decreasing the microscopic bending rigidity $\kappa_0$ (or increasing the microscopic Young's modulus $Y_0$) can be viewed as a proxy for increasing the temperature in experiments on graphene ribbons.) In all simulation runs the initial configuration was chosen to be planar. A typical run consisted of up to $2\cdot10^{9}$ time steps, or $10^{7}\tau$, where $\tau=a\sqrt{m/k_BT}$  is the reduced unit of time with $m=1$  being the the vertex mass and $k_BT$ ($k_B$ being the Boltzmann constant) setting the unit of energy. The step size was set to $5\cdot10^{-3}\tau$. Converted into execution time, each simulation takes between 24 and 60 hours on a single NVIDIA GTX 790Ti Graphical Processing Unit (GPU).

\section{Results}
\label{sec:results}

In order to demonstrate that the coarse-grained model presented in the previous section can indeed capture the long time- and length-scale behavior of graphene sheets, we first studied the spectrum of height fluctuations $h({\bf x})$ of rectangular sheets of size $100 \times 86.6$. As in previous atomistic Monte Carlo simulations of graphene~\cite{los09,zakharchenko10} we used periodic boundary conditions, where the bounding box was allowed to change its size, while maintaining zero external stress. This is achieved by running simulations in the NPT ensemble~\cite{frenkelB}. Theory~\cite{aronovitz88,guitter89,ledoussal92,nelsonB} predicts that the height fluctuations in momentum space scale as
\begin{equation}
\left<\left|h({\bf q})\right|^2 \right> = \frac{k_B T}{A \kappa_R(q) q^4},
\label{eq:height_fluctuations_sheet}
\end{equation}
where $A$ is the sheet area,  $h\left({\bf q}\right) = \int \left(d^2{\bf x}/A\right)\, e^{-i {\bf q} \cdot {\bf x}}\, h\left({\bf x}\right)$, and the renormalized bending rigidity scales as 
\begin{eqnarray}
\label{eq:renormalized_kappa_q}
\kappa_R(q) & \sim & \left\{
\begin{array}{c c}
\kappa_0, & q \gg q_\textrm{th}\\
\ \ \kappa_0 \left(q/q_\textrm{th}\right)^{-\eta}, \ \ & q \ll q_\textrm{th}
\end{array}
\right. .
\end{eqnarray}
The scaling exponent $\eta\approx 0.80-0.85$ quantifies the scale dependence of the bending rigidity driven by thermal fluctuations in the range of wavevector up to the transition scale $q_\textrm{th}$ above which thermal fluctuations are no longer significant:  
\begin{equation}
q_\textrm{th}=\sqrt{\frac{3 k_B T Y_0}{16 \pi \kappa_0^2}}.
\label{eq:q_thermal}
\end{equation}
For graphene at room temperature the transition wavevector is $q_\textrm{th} \approx 0.16 \textrm{\AA}^{-1}$.

\begin{figure}[t!]
\begin{centering}
\includegraphics[scale=0.52]{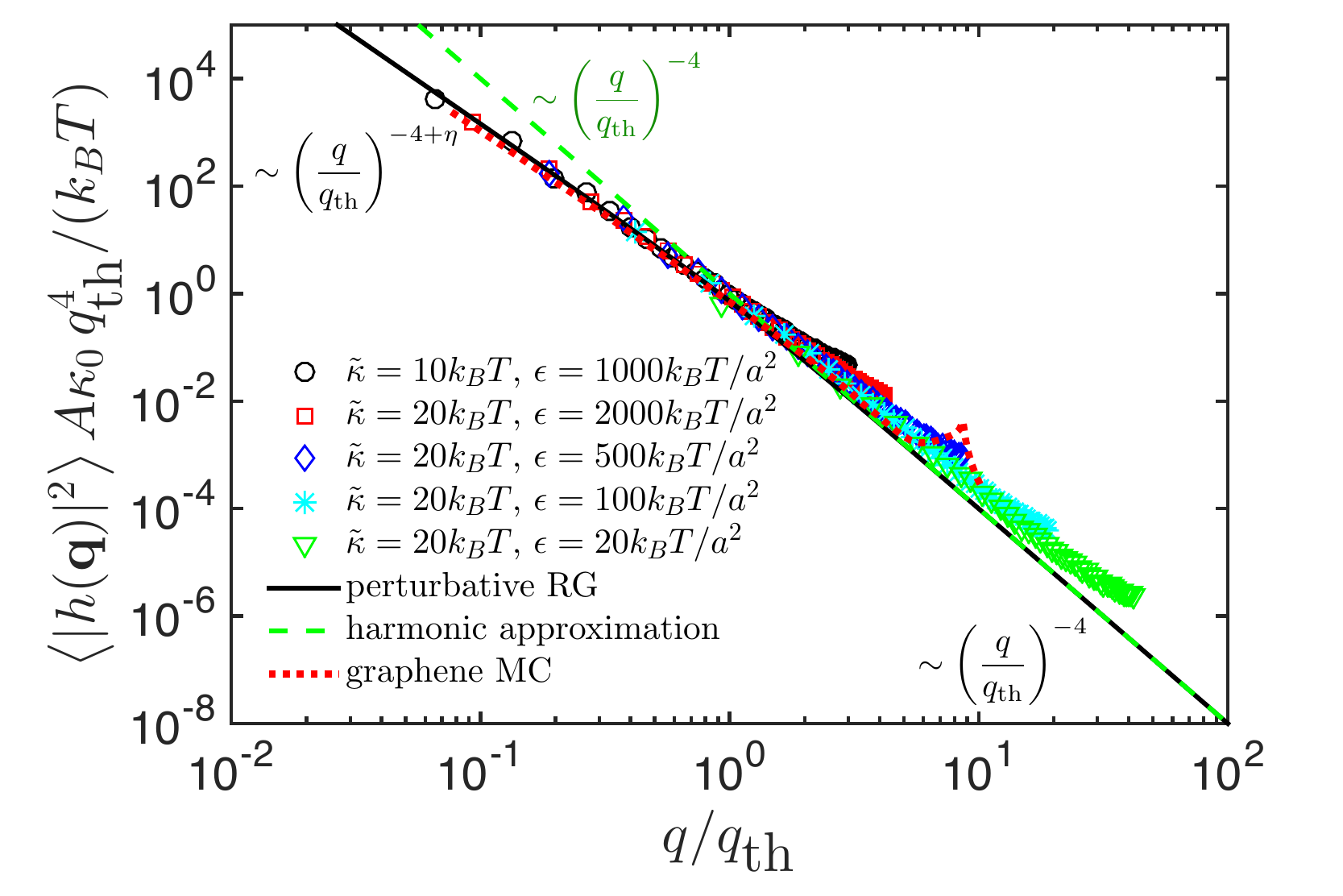}
\par\end{centering}
\caption{(Color online) Scaling collapse for height fluctuations $\left<\left|h({\bf q})\right|^2\right>$ of a rectangular sheets of size $100 \times 86.6a^2$ characterized with different values of bending rigidity $\tilde \kappa$ and spring constant $\epsilon$ that are defined in Eqs.~(\ref{eq:dihedral_bending}) and (\ref{eq:stretching}). For large wavevectors $q \gg q_\textrm{th}$ height fluctuations are well described with the harmonic approximation (dashed green line), where the renormalized bending rigidity $\kappa_R(q)$ can be approximated with the bare bending rigidity $\kappa_0$ [see Eq.~(\ref{eq:renormalized_kappa_q})]. For small  wavevectors $q \ll q_\textrm{th}$ the renormalization of bending rigidity $\kappa_R(q)$ with the characteristic exponent $\eta$ becomes apparent. Solid black line corresponds to the perturbative renormalization group result from Ref.~\cite{kosmrlj16}. The dotted red line is adapted from the atomistic Monte Carlo simulations in Ref.~\cite{zakharchenko10} and the small bump at $q/q_\textrm{th}\approx 5-10$ corresponds to wavevectors close to the edge of the first Brillouin zone.
}
\label{fig:height_correlations}
\end{figure}

Figure~\ref{fig:height_correlations} shows that both the atomistic Monte Carlo simulations~\cite{los09,zakharchenko10} and our coarse grained simulations agree quite well with the predicted scalings in Eqs.~(\ref{eq:height_fluctuations_sheet}) and ~(\ref{eq:renormalized_kappa_q}) as long as the wavevectors $q$ are much smaller than the microscopic cutoff $\Lambda~\sim 1/a$. In atomistic simulations the microscopic length cutoff is related to the characteristic distance $d$ between nearest-neighbor carbon atoms, while in our coarse-grained approach it is related to the lattice constant $a$.

We then explored graphene-like ribbons in a wide range of aspect ratios with $W=10\mbox{--}40 a$ and $L=40\mbox{--}300a$, where the strip is able both to flap as well as to twist along the long ($L$) direction. Slow elastic modes in the system make both reaching the thermal equilibrium and collecting statically independent samples for computing thermal averages a challenge. Typical autocorrelation times were around  approximately $10^4\tau$, which allowed for sampling the MD trajectories at time intervals $\Delta \tau = 2.5\cdot10^4\tau$ in order to obtain $\approx200\mbox{--}400$ statistically independent samples, resulting in the typical error of $\lesssim5\%$ for measured quantities.

\begin{figure}[tb]
\begin{centering}
\includegraphics[scale=0.5]{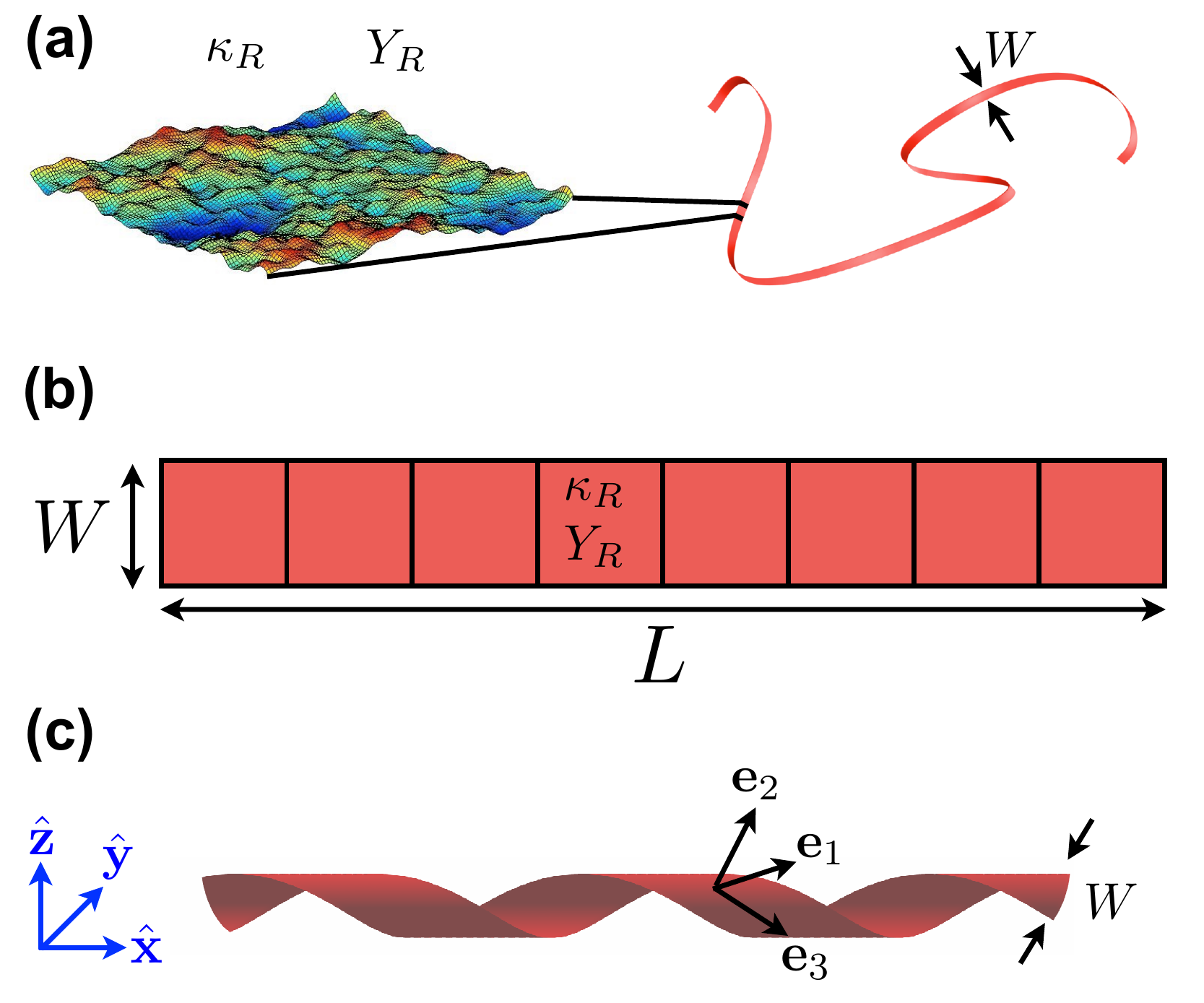}
\par\end{centering}
\caption{(Color online) Renormalization of ribbon elastic constants due to thermal fluctuations. (a) Snapshot of thermal fluctuations on a $W\times W$ patch of ribbon, the latter represented schematically, where color encodes the height of fluctuations with red (blue) describing positive (negative) height fluctuations. The effect of thermal fluctuations is to renormalize the bending rigidity $\kappa_R(W)$ and the Young's modulus $Y_R(W)$ on the scale of ribbon width $W$ according to Eqs.~(\ref{eq:renormalized_elastic_constants}).
(b) Coarse-grained ribbon is constructed with square blocks of size $W\times W$ with the renormalized elastic constants $\kappa_R(W)$ and $Y_R(W)$.
(c) A coarse-grained representation of ribbon configurations tracks the orientation of an orthonormal triad $\left(\hat{e}_1(s),\hat{e}_2(s),\hat{e}_3(s)\right)$
relative to a lab-frame $\left(\hat{e}_x,\hat{e}_y,\hat{e}_z\right)$ as a function of arclength $s$ along the ribbon~\cite{kosmrlj16}.
\label{fig:renormalization_sketch}}
\end{figure}

Before we discuss MD simulations, we briefly summarize the theoretical study in Ref.~\cite{kosmrlj16}. Here, renormalization group methods were used to demonstrate that ribbons behave like highly anisotropic polymers, with however strongly renormalized  
width-dependent elastic constants. A heuristic understanding arises from coarse-graining and constructing a ribbon with $L/W \gg 1$ into square membrane blocks of size $W \times W$ (see Fig.~\ref{fig:renormalization_sketch}). Thermal fluctuations generate a width-dependent bending rigidity $\kappa_R(W)$ and Young's modulus $Y_R(W)$ according to \cite{kosmrlj16}
\begin{subequations}
\label{eq:renormalized_elastic_constants}
\begin{eqnarray}
\label{eq:renormalized_kappa}
\kappa_R(W) & \sim & \left\{
\begin{array}{c c}
\kappa_0, & W \ll \ell_\textrm{th}\\
\ \ \kappa_0 (W/\ell_\textrm{th})^\eta, \ \ & W \gg \ell_\textrm{th}\label{eq:kappa_r}
\end{array}
\right.,\\
Y_R(W) & \sim & \left\{
\begin{array}{c c}
Y_0, & W \ll \ell_\textrm{th}\\
Y_0 (W/\ell_\textrm{th})^{-\eta_u}, & W \gg \ell_\textrm{th}
\end{array}
\right.,
\end{eqnarray}
\end{subequations}
with exponents $\eta\approx 0.80-085$ and $\eta_u=2-2\eta\approx 0.30-0.40$ characteristic of thermalized sheets~\cite{aronovitz88,guitter89,ledoussal92}. Evidently, the renormalization only becomes important for ribbons  whose width $W$ is larger than the thermal length scale discussed in the Introduction. A more precise estimate of $\ell_\textrm{th}$ is \cite{kosmrlj16}
\begin{equation}
\ell_\textrm{th}\equiv \frac{\pi}{q_\textrm{th}}= \sqrt{\frac{16 \pi^3 \kappa_0^2}{3 k_B T Y_0}},
\end{equation}
where $q_\textrm{th}$ is defined in Eq.~(\ref{eq:q_thermal}).
As discussed above, this scale is on the order of nanometers at room temperature for graphene and related atomically thin covalently-bonded sheets, indicating very large thermal renormalization even for ribbons with relatively modest widths $W\gg\ell_\textrm{th}$. 
If we characterize the coarse-grained orientations of the ribbon by rotations of the orthonormal triad $\left[\hat{e}_1(s),\hat{e}_2(s),\hat{e}_3(s)\right]$ (see Fig.~\ref{fig:renormalization_sketch}c) as a function of arc length $s$ along the ribbon, $d\hat{e}_i/ds = \vec\Omega \times \hat{e}_i$, the ribbon free energy takes the form \cite{landauB}

\begin{equation}
\label{eq:ribbon}
F= \frac{1}{2} \int_0^L \!ds \,\left[ A_1 {\Omega_1}^2 + A_2{\Omega_2}^2 + C{\Omega_3}^2 \right].
\end{equation}
The renormalized one dimensional ribbon bending rigidities $A_1$, $A_2$ and twisting rigidity $C$ are also strongly width-dependent \cite{kosmrlj16} 
\begin{subequations}
\begin{eqnarray}
A_1 & \sim & W \kappa_R(W),\\
A_2 & \sim & W^3 Y_R(W), \\
C & \sim & W \kappa_R(W).
\end{eqnarray}
\end{subequations}
Note that $A_2 \gg A_1, C$ for ribbons with large F\"oppl-von K{\'a}rm{\'a}n numbers $\textrm{vK}\sim\left(W/h\right)^2$. Ribbons thus behave like highly anisotropic polymers with persistence length~\cite{panyukov00}
\begin{equation}
\label{eq:persistence_length}
\ell_p = \frac{2}{k_B T \left(A_1^{-1}+A_2^{-1}\right)} \approx \frac{2 W \kappa_R(W)}{k_B T},
\end{equation}
where the $W-$dependence of $\kappa_R\left(W\right)$ in Eq.~(\ref{eq:kappa_r}) indicates a strong breakdown of Hookean elastic theory. 

\begin{figure}[t!]
\begin{centering}
\includegraphics[scale=0.5]{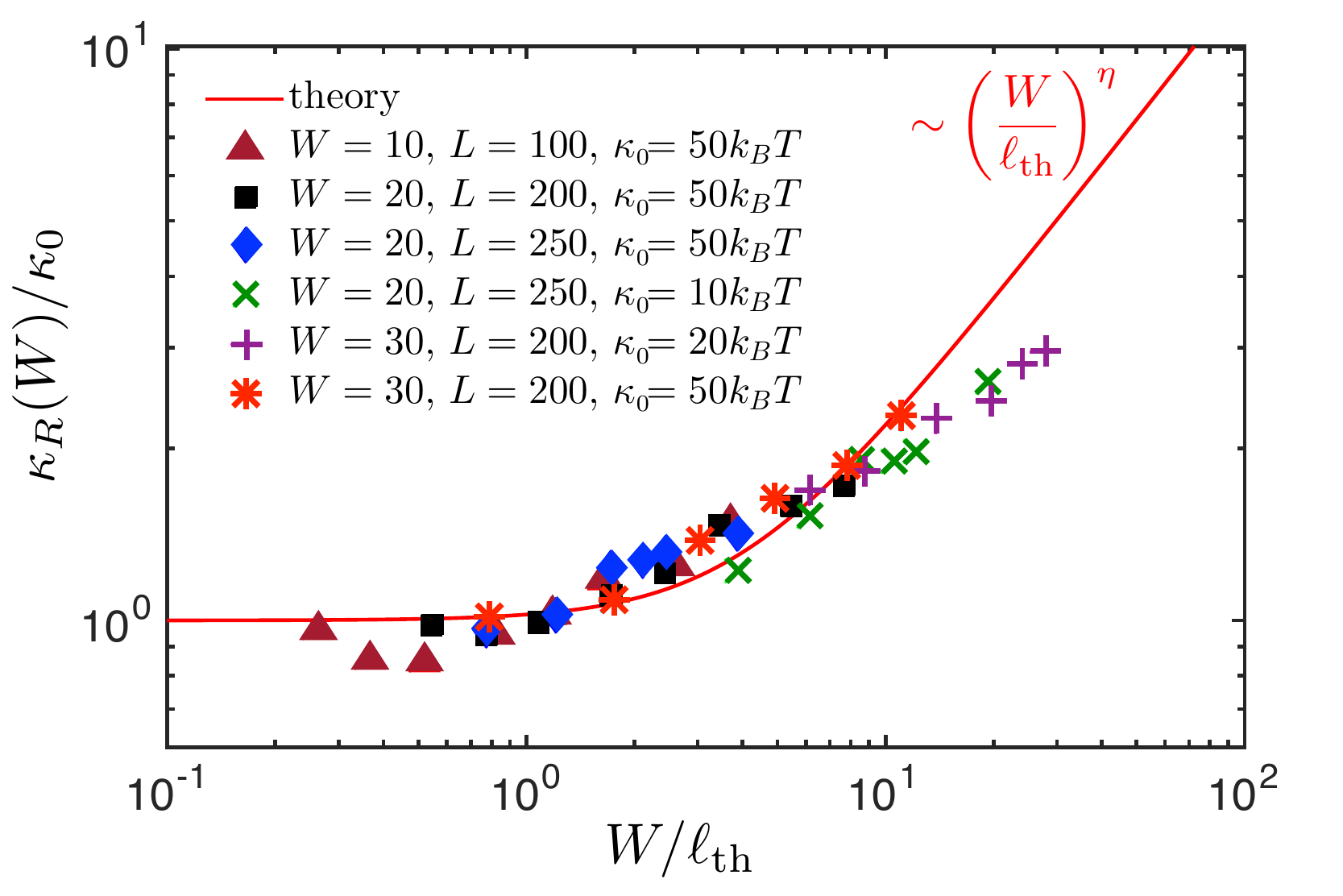}
\par\end{centering}
\caption{(Color online) Renormalized bending rigidity $\kappa_R(W)$ for ribbons of width $W$ extracted from simulation measurements of the persistence length $\ell_p$ [see Eq.~(\ref{eq:persistence_length})]. The scaling collapse for ribbons of various dimensions and bare bending rigidities $\kappa_0$ is consistent with  Eq.~(\ref{eq:renormalized_kappa}), where the red theoretical curve was obtained with renormalization group procedure described in Ref.~\cite{kosmrlj16}. The theoretical calculations were done with periodic boundary conditions across the ribbon width. We attempted to account for our different boundary conditions (clamped at one end and free at other) by shifting the theoretical curve along the horizontal direction. In these simulations, the value of ratio $W/\ell_\textrm{th}$ was set by tuning the Young's modulus $Y_0$ of the ribbon.}
\label{fig:renormalized_kappa}
\end{figure}

To test the theoretical predictions above against our simulations, the persistence length $\ell_p$ was determined from the decay of the autocorrelation function of the tangent vectors $\vec{t}(s) \equiv \vec{e}_3(s)$ to the midline along the ribbon's length \cite{panyukov00}
\begin{equation}
\left<{\bf t}\left(s\right) \cdot {\bf t}\left(s+x\right)\right> = e^{-x/\ell_p}.
\end{equation}
Here, the averaging was done over all possible pairs of tangent vectors that were separated by distance $x$ along the ribbon backbone and also over all 200 independent ribbon configurations. Measured persistence lengths $\ell_p$ were then used to obtain the values of renormalized bending rigidities as $\kappa_R\left(W\right) = k_B T \ell_p/\left(2 W\right)$ [see Eq.~(\ref{eq:persistence_length})], which are displayed in Fig.~\ref{fig:renormalized_kappa}. When $W/\ell_\textrm{th}\lesssim1$, the renormalized elastic constant is approximately independent of $W$, consistent with classical elasticity theory. However, the resulting data collapse is consistent with  a scale-dependent renormalized bending rigidity that starts increasing for ribbons whose width $W$ is larger than the thermal length scale $\ell_\textrm{th}$, similar to Eq.~(\ref{eq:renormalized_kappa}). More extensive simulations would be needed to convincingly demonstrate the $\left(W/\ell_\textrm{th}\right)^\eta$ scaling of $\kappa_R/\kappa$ expected for $W/\ell_\textrm{th}\gg1$. Note that our extensive computer simulations are unfortunately limited to $W/\ell_\textrm{th}\lesssim30$, as opposed to the values $5\lesssim W/\ell_\textrm{th}\lesssim 5000$ accessible in thermalized graphene ribbons at room temperature \cite{blees15}.

\begin{figure}[t]
\begin{centering}
\includegraphics[scale=0.5]{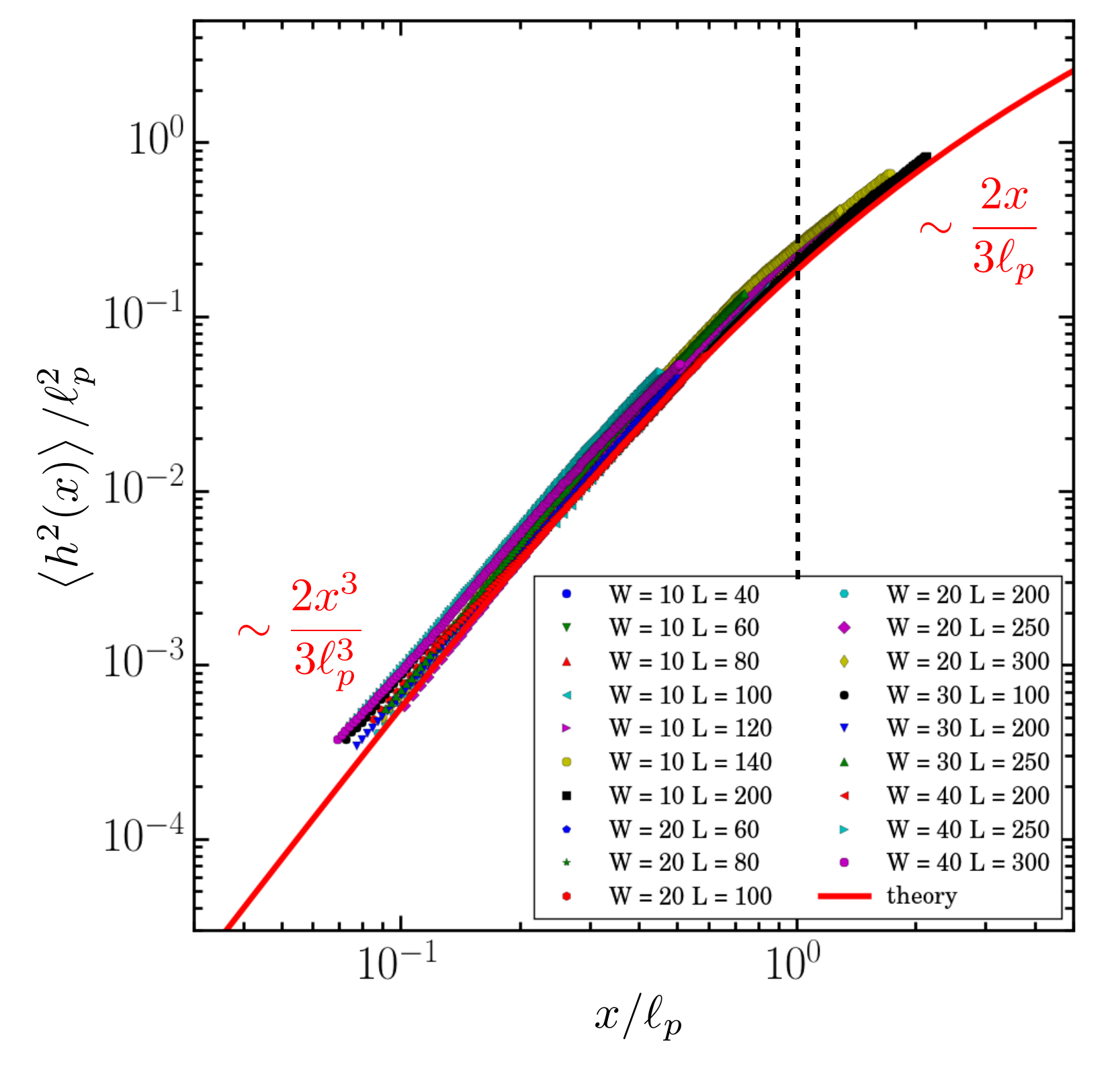}
\par\end{centering}
\caption{
(Color online) Scaling collapse for ribbon height fluctuations $\left\langle h^{2}\left(x\right)\right\rangle /\ell_{p}^{2}$, where $x$ represents the distance from the clamped end along the ribbon backbone. The red line indicates analytically predicted curve for anisotropic ribbons in Eq.~(\ref{eq:height_fluctuations}), which is in a good agreement with results of numerical simulations with no adjustable parameters. The persistence length $\ell_{p}$ is measured from the simulations by computing the autocorrelation function of tangent vectors to the ribbon's midline as described in text.  
\label{fig:lp_scaling}}
\end{figure}

Finally, we test whether long fluctuating ribbons in fact behave like anisotropic polymers by measuring the ribbon height fluctuations $\langle h^{2}\left(x\right)\rangle$, where $h\left(x\right)$ is the deviation away from the average ribbon position, and $x$ represents the distance from the clamped end along the ribbon backbone. For each value of $x$ at a given time $\tau$ the value of $h\left(x\right)$ was determined by averaging over the width of the ribbon, i.e., $h\left(x\right)=\frac{1}{M_x}\sum_{i}h\left(x,y_{i}\right)$, where $i$ counts all vertices at distance $x$ from the clamp and $M_x$ is the total number of such vertices. In order to ensure that results were not affected by this averaging procedure, the data was also analyzed by extracting the central bisecting the width of the ribbon and computing the same root mean-square average of $h\left(x\right)$. Results for the two approaches are nearly identical (data not shown).
In Fig.~\ref{fig:lp_scaling} we show the scaling collapse of measured height fluctuations $\left\langle h^{2}\left(x\right)\right\rangle$ using our measured persistence length $\ell_p$, in excellent agreement with height fluctuations for anisotropic polymers~\cite{kosmrlj16}
\begin{equation}
\label{eq:height_fluctuations}
\left<h^2(x)\right> = \left\{
\begin{array}{c c}
2 x^3/(3 \ell_p), & x \ll \ell_p \\
2 x \ell_p/3, & x \gg \ell_p
\end{array}
\right..
\end{equation}
Close to the clamp ($x\ll \ell_p$) ribbons behave like stiff cantilevers, with an $x^3$ dependence for the mean square height fluctuation, while far away ($x\gg \ell_p$) they transition to random walk behavior. Because we do not include distant self-avoidance in our simulations, we have $\sqrt{\left<h^2\left(x\right)\right>} \sim x^{1/2}$, in this regime, rather than the behavior of a self-avoiding random walk.

\section{Conclusions}
\label{sec:conclusions}

In the past theoretical studies and simulations have primarily focused on the effects of thermal fluctuations in flat sheets. The recent experimental realization of graphene kirigami~\cite{blees15}, however,  and the possibility of growing graphene on curved substrates via chemical vapor deposition, motivates an analysis of the role of geometry. With the coarse-grained molecular dynamics simulations of ribbons in this work and with the accompanying theoretical study in Ref.~\cite{kosmrlj16},  we have demonstrated that long ribbons behave like interesting hybrids between flat sheets and asymmetric polymers. Just like polymers, ribbons become semi-flexible beyond a characteristic persistence length $\ell_p$. The persistence length, however, scales non-trivially ($\ell_p \propto W^{1+\eta}/T^{1-\eta/2}$) with the ribbon width $W$ and with temperature $T$, when the thermal length scale becomes smaller than the ribbon width ($\ell_\textrm{th} \lesssim W$). This is a direct consequence of the renormalization of the ribbon bending rigidity at the scale of the ribbon width.

The spontaneous curvature of sheets also leads to new surprising phenomena as was demonstrated in recent Monte Carlo simulations~\cite{paulose12} and in a theoretical study~\cite{kosmrlj17} of thermalized spherical shells. In spherical shells thermal fluctuations produce effective negative surface tension, which can be interpreted as an effective external pressure. As a result thermal fluctuations reduce the critical buckling pressure for spherical shells up to a point, that shells, which are larger than some temperature dependent critical radius, become crushed even when the pressure difference between the inside and outside of the shell is zero~\cite{kosmrlj17}. A similar result was observed in numerical simulations of carbon nanotubes~\cite{zhang06}, where thermal fluctuations reduced the critical axial load.

Whilst the essential electronic properties of pure graphene are barely affected by shape fluctuations we see that its mechanical moduli, typified by the bending rigidity, are strongly length-scale dependent and tunable geometrically. This  opens the door to designable elements for metamaterials with targeted mechanical properties whilst retaining all the other material virtues of pure graphene.

\begin{acknowledgments}
R.S. would like to acknowledge the financial support from EPSRC via grant EP/M009599/1 and from BBSRC via grant BB/N009789/1. Support for M.J.B.~and D.R.N.~by the National Science Foundation, through the NSF DMREF program via grant DMR-1435794 and DMR-1435999, is gratefully acknowledged. Support from the NSF via grant DMR-1306367 and through the Harvard Materials Research and Engineering Center through Grant DMR1420570 is acknowledged by A.K. and D.R.N. 
\end{acknowledgments}


%

\end{document}